\documentclass[superscriptaddress,aps,prb,twocolumn,groupaddress]{revtex4}
\usepackage{amssymb}
\usepackage{latexsym}
\usepackage{amsmath}
\usepackage[mathcal]{eucal}
\usepackage{graphicx}
\usepackage{dcolumn}
\usepackage{amsfonts}
\usepackage{bm}
\usepackage{epsfig}
\usepackage{array}

\newcommand{\be}{\begin{equation}}
\newcommand{\ee}{\end{equation}}
\newcommand{\bea}{\begin{eqnarray}}
\newcommand{\eea}{\end{eqnarray}}
\newcommand{\bal}{\begin{align}}
\newcommand{\eal}{\end{align}}

\renewcommand{\vec}[1]{{\bm #1}}

\bibpunct{[}{]}{,\!}{n}{,}{,} 

\begin{document}
\title{Vertical loop nodes in iron-based superconductors}
\author{M. Khodas}
\affiliation{Department of Physics and Astronomy, University of Iowa, Iowa City, Iowa 52242, USA}
\author{A. V. Chubukov}
\affiliation{Department of Physics, University of Wisconsin, Madison, Wisconsin 53706, USA}

\begin{abstract}
We consider Fe-based superconductors with  $s^{+-}$ gap with accidental nodes on electron pockets.
 We analyze how the gap structure changes if we include into the
  consideration the hybridization between the two electron pockets (the inter-pocket hopping term with momentum $(\pi,\pi,\pi)$.
  We derive the hybridization term and relate it to the absence  of inversion symmetry in the Fe-plane
  because of two non-equivalent locations of pnictogen (chalcogen) above and below the plane.
   We find that the hybridization tends to eliminate the nodes -- as it increases, the pairs of neighboring nodes approach each other,
   merge and disappear once the hybridization  exceeds a certain  threshold.
   The nodes disappear first around $k_z =\pi/2$, and vertical line nodes split
    into  two vertical loops  centered at $k_z =0$ and
      $k_z = \pi$.
 We also show that  the hybridization moves
  the nodes along the loops away from the normal state Fermi surfaces.
  This creates a subset of $k-$points at which the peak in the spectral function
  does not shift as the system enters into a superconducting state (``no-shift'' lines). These ``no-shift'' lines evolve with increasing hybridization in highly non-trivial manner and eventually form horizontal  loops in $(k_x, k_y)$ plane, surrounding the  nodes.
    Both vertical line nodes and horizontal ``no-shift'' loops surrounding them should be detectable in photoemission experiments.
 \end{abstract}

\maketitle

\section{Introduction}

Understanding high-temperature superconductivity in doped Fe-pnictides and
Fe-chalcogenides remains
 the top priority for the condensed matter community~\cite{review,review_1,review_2,review_3}.
 Superconductivity in weakly/moderately doped systems is  generally  believed to be
  the consequence of the  complex geometry of the Fermi surface (FS),
   which consists of hole and electron pockets located in
   different regions of the Brillouin zone. The prevailing scenario is that the
     superconducting
     gap has an $s-$wave symmetry,
      but
       changes sign  between hole and electron pockets~\cite{mazin_1,kuroki_1}, and may even have accidental nodes~\cite{review,cvv}.

Previous   studies of the pairing mechanism and the gap structure in Fe-based superconductors mostly focused on a 2D model of a single Fe plane with adjacent pnictogen/chalcogen atoms located immediately above and below it~\cite{review,kuroki_1,cao,10_orbital,10_orbital_1,hu_hao,leni,daghofer}.
Electrons from 3d-orbitals of Fe hop mostly indirectly, via 4p pnictogen/chalcogen sites.
The low-energy structure around $E_F$ obtained from a
 3d-4p hopping has been fitted~\cite{review,kuroki_1,cao,hu_hao,leni,daghofer} to an effective ``Fe only'' 2D 5-orbital tight-binding Hamiltonian with inter-orbital and intra-orbital hopping terms.
The band structure and the location of the electron and hole FSs have been deduced by evaluating the eigenvalues of the 5-orbital tight-binding Hamiltonian and analyzing the energy profile.
The pairing problem can be most straightforwardly analyzed in the band basis,
by  associating the operators corresponding to eigenfunctions with band operators and re-writing the interaction in the band basis~\cite{review}.
The band operators are  linear combinations of orbital operators, and the interactions in the band basis are the ones in the orbital basis, dressed by ``coherence factors'' associated with the fact that each eigenfunction is a  linear combinations of Fe-orbitals.
The dressed interactions inherit angular dependencies from the  coherence factors.
Solving for gaps on different FSs, one then generally obtains angular-dependent gaps, even in the s-wave case.
By generic reasoning, the $s-$wave
gaps on hole pockets contain $\cos 4 n \phi$ harmonics ($n =0,1,2..$) where $\phi$ is the angle along the hole pocket, while the gaps on electron pockets contain both $\cos 4 n \theta $ and $\cos 2\theta (2n+1)$ harmonics,
with the angle $\theta$ counted relative to,
 e.g., $x$ axis for one pocket and $y$ axis for the other~\cite{cvv,maier,maiti}.
 Numerical analysis shows~\cite{review,maiti} that $\cos 2\theta$ harmonic of the gap is the strongest one, and in some materials it is large enough to induce accidental nodes on electron FSs.

The actual BZ, however, contains two Fe atoms because pnictogen/chalcogen is located below and above Fe plane in checkerboard order.
The positions of pnictogen/chalcogen atoms are shifted by half of lattice spacing in both $x$ and $y$ directions relative to the positions of Fe.
 A half of pnictogen/chalcogen
 atoms are located above Fe plane and half are located below Fe plane.
 As a consequence, for half of Fe atoms hoping to pnictogen/chalcogen means hopping down along $z$ and for the half it means hopping up along $z$.
 Therefore, the symmetry of the original lattice is lower than the symmetry of the Fe lattice, and the effective tight-binding Hamiltonian should generally contain two type of terms -- the ones with zero momentum transfer, and the ones with momentum transfer $\vec{Q} = (\pi, \pi)$~\cite{hybridization,mazin} (here and below we set interatomic spacing to one).

The tight-binding Hamiltonian considered in earlier studies~\cite{review} includes only the terms with zero momentum transfer.
For such a Hamiltonian, the transformation from  1-Fe  BZ to 2-Fe BZ  is just a rotation in a  momentum space: momenta $k_x$ and $k_y$ from 1-Fe zone are transformed into ${\tilde k}_a = k_x + k_y$ and  ${\tilde k}_b = k_x - k_y$ in the 2-Fe zone.
The gap structure is obviously not affected by this transformation, up to a change of variables.
In particular, if the gap has accidental nodes in 1-Fe zone, it will have accidental nodes in 2-Fe zone as well.

\begin{figure}[h]
\begin{center}
\includegraphics[width=1.0\columnwidth]{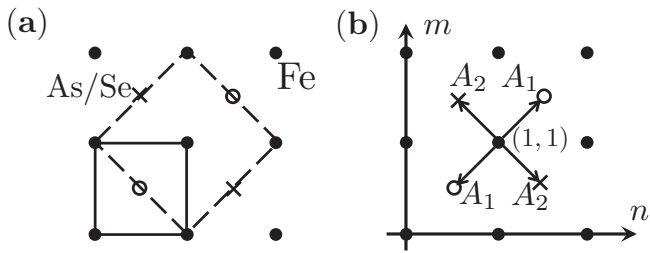}
\caption{(a) Schematic representation of a single layer of pnictide/chalcogenide.
The solid square designates a Fe (solid circles) only lattice unit cell.
The As/Se atoms form two inequivalent subsets.
The atoms denoted by a cross and empty circles are shifted off the Fe
 plane in opposite directions.
(b) The hopping amplitudes to a neighboring pnictogen/chalcogen site are generally different for pnictigen/chalcogen above and below Fe plane, i.e., in general $A_1 \neq A_2$.
}
\label{fig:layer}
\end{center}
\end{figure}

In this paper we demonstrate that the effective tight-binding Hamiltonian also contains hopping terms with momentum transfer $\vec{Q}$,  and that these terms affect the physics in a quantitative way, and, in particular,  substantially modify the gap structure.
The $(\pi,\pi)$ term vanishes if one approximates a 2D 5-orbital model by just $d_{xz}$ and $d_{yz}$ orbitals, as has been done in some earlier studies~\cite{raghu,hu_hao} but is present once one adds a $d_{xy}$ orbital (the $(\pi,\pi)$ term describes the hopping from $d_{xy}$ orbital to $d_{xz}$ or $d_{yz}$ orbital).
Because the low-energy hole and electron states in the full 2D tight-binding Hamiltonian have contributions from $d_{xz}, d_{yz}$, {\it and} $d_{xy}$ orbitals~\cite{review,kuroki_1,cao,10_orbital,10_orbital_1,leni,daghofer,amalia}, the $(\pi,\pi)$ hopping term survives the transformation to band formalism and  becomes $\psi^\dagger_{\vec{k}} \psi_{\vec{k}+ \vec{Q}}$, where $\psi$ are band operators.
As the two electron FSs are separated exactly by $\vec{Q}$, such term gives rise to a hybridization between the two electron pockets.
The new band operators for low-energy fermions then become linear combination of the original $\psi_{\vec{k}}$, and the pairing interaction, re-expressed in terms of the new band operators,  acquires additional coherence factors, which modify its angular dependence.
The modification of the angular dependence of the interaction  in turn modifies the angular dependence of the  gaps on electron FSs.
This physics must be present in 10-orbital studies~\cite{10_orbital,10_orbital_1,amalia}, which treat orbital excitations with momentum $\vec{k}$
and $\vec{k} +\vec{Q}$ as separate degrees of freedom, but we are not aware of the attempt to fit the results of 10-orbital numerical studies by an effective tight-binding Hamiltonian with an additional $(\pi,\pi)$ term.

In the previous work~\cite{khodas_1} we demonstrated that hybridization plays a crucial role in systems with only electron pockets, like KFe$_2$Se$_2$.
Several groups analyzed the pairing problem in KFe$_2$Se$_2$ in the absence of hybridization.
If the interaction between the two electron pockets is repulsive, the pairing is possible if inter-pocket interaction exceeds intra-pocket repulsion, but the gaps on the two electron pockets should have opposite signs~\cite{kfese,maiti,comm_1}. Such a state has a $d-$wave symmetry because the gap changes sign under rotation by $90^{\circ}$.
The hybridization mixes the two original electron pockets into two new electron pockets, each contains states from both original pockets.
The two new pockets split upon hybridization, and the solution of the pairing problem for strong enough hybridization shows that the system prefers to form an $s-$wave gap with opposite sign on the two hybridized FSs~\cite{mazin,khodas_1}. In terms of original fermions the condensate wave function for such an s-wave state is made out of fermions from different electron FSs, i.e., a pair has a momentum $\vec{Q}$.
In terms of new, hybridized fermions, such a state is a conventional one, with zero total momentum of a pair.
The microscopic analysis of the pairing in the presence of hybridization  shows~\cite{khodas_1} that, with increasing hybridization, the system undergoes a two-step transition from a $d-$wave state to a $d \pm i s$  state and then to an  $s^{+-}$ state.

In this paper we analyze the role of hybridization in systems in which both hole and electron pockets are present and the gap has $s^{+-}$ form.
 We take the gap structure from the 2D tight-binding Hamiltonian without $\vec{Q}$ terms as an input and analyze how it changes once we add additional hopping with momentum transfer $\vec{Q}$.
At first glance, the hybridization between the two electron pockets should not lead to qualitative changes in the $s^{+-}$ gap because the
gaps
order parameter on these pockets on average
have the same sign.
We show, however, that the actual situation is more tricky and hybridization does give rise to {\it qualitative} changes in the gap structure in systems like  BaFe$_2$(As$_{1-x}$P$_x$)$_2$ (Ref.~\cite{matsuda}), LaOFeP (Ref. \cite{lafop}) and LiFeP~(Ref.~\cite{lifep}), in which $s^{+-}$ gap has accidental nodes, which most likely reside on electron pockets.
By symmetry, there are four nodes on each electron pocket (8 nodes in total).
We show that the hybridization brings the neighboring nodes  close to each other. When the  hybridization reaches a certain threshold, the nodes coalesce and disappear (see Figs.~\ref{fig:1},\ref{fig:3}).

The importance of hybridization between electron pockets was recognized in Ref.~\onlinecite{10_orbital_1}. The authors of that work conjectured that it may affect the
location of the gap nodes. Our explicit calculations do show that the nodal structure is indeed affected by hybridization.

We follow the evolution of the nodes before the hybridization reaches a threshold.
We show that at any non-zero hybridization the nodes do not reside on the normal state FSs but are shifted into the $k-$region between the hybridized FSs, where the ARPES intensity above $T_c$ is peaked  at a finite negative frequency.
As a result, the peak in the ARPES spectra at the nodal point doesn't stay intact, as in a ``conventional'' superconductor with gap nodes, but shifts towards a smaller frequency below $T_c$.
We further show that each nodal point is surrounded by ``no-shift" lines (NSL), at which quasiparticle energies in the normal and superconducting states are equal, i.e., the peak in the ARPES spectrum does not shift when the system becomes a superconductor (Figs.~\ref{fig:2},\ref{fig:3}).
Without hybridization, these NSL are radial beams transverse to the FS, in the direction along which the gap vanishes.
Upon hybridization, NSL rapidly evolve, rotate by $90^{\circ}$, and
transform into loops directed {\it along} the FSs.
Inside the loop, ARPES peak moves to a lower negative frequency below $T_c$, outside the loop it moves to a higher negative frequency.
We propose to search for such NSL in ARPES measurements on BaFe$_2$(As$_{1-x}$P$_x$)$_2$ and other $s^{+-}$ superconductors with gap nodes, like LaOFeP~\cite{lafop} and LiFeP~\cite{lifep}.

We next consider the effect of hybridization in 3D systems.
For 1111 systems with a simple tetragonal lattice, the hybridization term has 3D momentum $(\pi,\pi,0)$, 3D effects are secondary, and the consideration is qualitatively the same as in 2D systems.
In 122 systems  with body-centered tetragonal lattice, the situation is more interesting.
We show that in these systems the hybridization between electron pockets
is three-dimensional, with hybridization vector $(\pi,\pi,\pi)$,
even if we neglect the $k_z$ dependence of the gap along the third direction within the 1Fe tight-binding Hamiltonian~\cite{3D}.
This is due to the fact that in 122 materials pnictogen/chalcogen atoms alternate in
checkerboard order between neighboring Fe planes, i.e., if for a given Fe plane
a pnictogen/chalcogen atom at a given $(x,y)$ is located above Fe plane, then in the neighboring Fe plane a pnictogen/chalcogen atom at the same $(x,y)$ is located below Fe plane~\cite{122}.
We show that, for such a structure, the hybridization term has 3D momentum $\vec{Q} = (\pi,\pi,\pi)$,
  and is the largest at $k_z = \pi/2$ (it actually  does not vanish, except for $k_z=0$ and $k_z = \pi$, even if we consider only $d_{xz}$ and $d_{yz}$ Fe orbitals).
We show that,
          at the same time,
          the critical value of the hybridization, at which the nodes disappear,
     is the smallest at $k_z = \pi/2$ and is much larger for $k_z =0$ and $k_z = \pi$.
The combination of the facts that at $\pi/2$ the value of the hybridization is the highest and the threshold value is the lowest implies that there must be a wide range of hybridizations when the nodes are eliminated near  $\pi/2$ but are still present near $k_z =0$ and $k_z =\pi$.
In this range vertical line nodes close up into vertical loop nodes centered at $k_z=0$ and $k_z = \pi$ (see Fig.~\ref{fig:4}).
The gap structure with such loop nodes has been proposed phenomenologically~\cite{matsuda} as the best candidate to fit the thermal conductivity data in BaFe$_2$(As$_{1-x}$P$_x$)$_2$, for which penetration depth, thermal conductivity, specific heat and NMR data all show~\cite{thermal_p} that the gap must have nodes, and at least  some ARPES results~\cite{laser_arpes} indicate that the nodes must be on the electron FSs.
Our result provides a microscopic explanation of vertical loop nodes.

The paper is organized as follows.
In the next section we present the model  of a $s^{+-}$ superconductor with nodes on electron pockets and introduce the hybridization amplitude.
In Sec. \ref{sec:2D} we discuss how the hybridization affects the nodes of the superconducting gap in systems with a simple tetragonal lattice structure.
We consider first a toy model with circular electron pockets and then consider a more realistic model of elliptical pockets.
We show that in both cases the neighboring nodes move towards each other as hybridization amplitude increases.
In Sec. \ref{sec:3D} we extend the analysis to a 3D body-centered tetragonal lattice and show that hybridization splits vertical line nodes into vertical loop nodes centered at $k_z =0$ and $k_z = \pi$.
In Sec.\ref{sec:exp} we discuss the experimental situation and comment on recent ARPES studies of the superconducting gap structure in BaFe$_2$(As$_{1-x}$P$_x$)$_2$.
We present our conclusions in Sec. \ref{sec:concl}.
We discuss technical issues in detail in three Appendices.
In  Appendix \ref{sec:hyb},  we  discuss in detail  the microscopic mechanism
of hybridization in 2D and 3D lattices and obtain explicit expressions for the hybridization amplitude for a simple tetragonal lattice structure and for a body-centered  tetragonal lattice structure.
In Appendix \ref{sec:app_A} we discuss the evolution of  NSL from radial beams in the direction transverse to the FS along which the gap vanishes to closed loops directed along the FSs.
In Appendix \ref{sec:app_B} we discuss subtleties in extracting the positions of gap nodes from ARPES data.

\section{The model}
\label{sec:model}

We consider  an $s^{+-}$ superconductor with two electron pockets at $(0,\pi)$ and $(\pi,0)$  in the 1FeBZ at
 any
given $k_z$, and the appropriate number (2 or 3) of  hole pockets centered at $(0,0)$.
 The actual pocket structure in Fe-based superconductors is somewhat more involved, i.e., low-energy states near $(0,\pi)$ and $(\pi,0)$ likely contain not only electron  pockets centered at these points, but also hole barrels,
 centered somewhat away from $(0,\pi)$ and $(\pi,0)$ (Ref.\cite{borisenko}).
These hole barrels will not play a role in our consideration and we neglect them.
In the absence of hybridization, the quadratic part of the Hamiltonian for electron pockets is
\begin{align}\label{H0}
H_2 =  \sum_{\vec{k}} \xi_{1,\vec{k}} \psi_{1,\vec{k}}^{\dag} \psi_{1,\vec{k}} + \xi_{2,\vec{k}+{\vec{Q}}} \psi_{2,\vec{k}+{\vec{Q}}}^{\dag} \psi_{2,\vec{k}+{\vec{Q}}}\, ,
\end{align}
where
$\psi_{1,2}$ describe two electron bands, one with low-energy excitations near ${\vec k}_1 = (0,\pi)$, and another with low-energy excitations near ${\vec k}_2 = (\pi,0) = {\vec k_1} + {\vec Q}$ (modulo $2\pi$), $\xi_{1,\vec{k}}$ and $\xi_{2,\vec{k}+{\vec{Q}}}$  are the corresponding  electron dispersions, and ${\vec Q} = (\pi,\pi)$.
  We  approximate fermion
excitations near the pockets by
$\xi_{1,\vec{k}} = v_F (\theta) (|{\vec k} -{\vec k}_1| -k_F( \theta)), ~~ \xi_{2,\vec{k}+{\vec{Q}}} = v_F (\theta+\pi/2) (|{\vec k} -{\vec k}_2|-k_F( \theta+\pi/2))$,
 where
 $\theta$ is the angle along each of the FSs counted from the $x-$axis~\cite{review,review_1,review_2,khodas_1}.  By virtue of tetragonal symmetry, $v_F (\theta) = v_F (1 +a \cos 2 \theta)$ and $k_F (\theta) = k_F (1 + b \cos 2 \theta)$. The parameter $b$ accounts for the eccentricity (ellipticity) of the FSs.  For
  1111 systems,  parameters $a$ and $b$ are
  essentially independent on $k_z$, and one can reduce the analysis to 2D model.
For 122 systems $a$ and $b$ do depend on $k_z$ and change sign and magnitude between $k_z =0$ and $k_z = \pi$.

We assume, following earlier works~\cite{review,review_1,review_2,review_3,kuroki_1}, that the dominant pairing interaction is between electrons and holes, and that its dependence on $k_z$ is weak and can be neglected.
We follow Refs.~\cite{cvv,maiti,maier} and approximate electron-hole interaction by  a constant term and by  $\cos 2 \theta$ term,
which changes sign between the two electron pockets.
Within this approximation, the gaps on hole FSs are angle-independent, while the gaps on the two electron FSs are $\Delta (1 \pm \alpha \cos 2 \theta_k)$.
The corresponding effective BCS Hamiltonian is
\begin{align}
\label{Hi}
H_{\mathrm{BCS}} = & \Delta \sum_k \left[( 1 - \alpha \cos 2 \theta_k ) \psi_{1,k} \psi_{1,-k}  + h.c \right] \nonumber \\
& +  \left[( 1 + \alpha \cos 2 \theta_k ) \psi_{2,k+Q} \psi_{2,-k-Q} + h.c. \right] \, .
\end{align}
We consider the case $\alpha >1$, when the gaps have accidental nodes in the absence of hybridization between the two electron pockets.
The hybridization term in the 1FeBZ has the form
\be\label{hyb}
H_{\vec{Q}} = \sum_{\vec{k}} \lambda(\vec{k}) \psi^{\dag}_{1\vec{k}} \psi_{2,\vec{k}+\vec{Q}}  + h.c.
\ee
We present the derivation of Eq.~(\ref{hyb}) in Appendix \ref{sec:hyb} for both 2D and 3D systems.
We show there that the prefactor $\lambda(\vec{k})$  is generally
 a complex number, which depends on both $k_z$ and $\vec{k}_{\perp}$.
 When only the $d_{xz}$ and $d_{yz}$ orbitals of Fe are considered, the hybridization vanishes completely in 2D case and  along particular directions in 3D case.
The $\vec{k}$-dependence becomes less strong once one adds into consideration $d_{xy}$ orbital along with spin-orbit interaction.
 In the latter case, $|\lambda(\vec{k})|$ does not vanish for any $\vec{k}$, although it is smaller at  $k_z=0$ and along $k_x = \pm k_y$ measured from the center of one of electron pockets.
The non-singular $\vec{k}$-dependence of $\lambda (\vec{k})$ is not essential to our analysis, and to simplify the presentation we approximate $\lambda (\vec{k})$ by a constant $\lambda$ for the rest of the paper.

Our goal is to analyze what happens with the nodes, and, more generally, with the fermion dispersion, when we solve for the pairing in the presence of the hybridization term.

\section{The effect of hybridization on the nodes, 2D case}
\label{sec:2D}

In this section we consider the case of a simple tetragonal lattice for which the pairing problem can be analyzed within a single  2D cross section (we recall that we treat interactions as independent on $k_z$).
In the presence of $H_{\vec{Q}}$, Eq.~\eqref{hyb}), the  quadratic part of the Hamiltonian for fermions near the two electron FSs becomes, instead of (\ref{H0})
\begin{align}\label{H01_1}
H_2 & \!=\!  \!\sum_{\vec{k}} \xi_{1,\vec{k}} f_{1,\vec{k}}^{\dag} f_{1,\vec{k}} +
\xi_{1,\vec{k}+ {\vec Q}} f_{2,\vec{k}+ {\vec Q}}^{\dag} f_{2,\vec{k}+ {\vec Q}}
\notag\\
&
+ \sum_{\vec{k}} \lambda \left[ f_{1,\vec{k}}^{\dag} f_{2,\vec{k}+ {\vec Q}} \!+\! h.c. \!\right]\, .
\end{align}
For convenience, we redefine ${\vec k}$
 and count it relative to ${\vec k}_1 = (0,\pi)$ for $f_1$ fermions and relative to ${\vec k}_2 = (\pi,0)$ for
 $f_2$ fermions, i.e.,
 absorb
${\vec Q}$ into new ${\vec k}$.
In these new notations, the quadratic Hamiltonian becomes
 \begin{align}\label{H01}
H_2 \!=\!  \sum_{j=1,2}\!\sum_{\vec{k}} \xi_{j,\vec{k}} f_{j,\vec{k}}^{\dag} f_{j,\vec{k}}
\!+\! \sum_{\vec{k}} \lambda \left[ f_{1,\vec{k}}^{\dag} f_{2,\vec{k}} \!+\! h.c. \right]\, .
\end{align}

To analyze the pairing, we now have to introduce new fermions which diagonalize the quadratic form in Eq.~\eqref{H01}, re-express the pairing interaction in terms of these new fermions, and solve for the gaps on electron FSs and
quasi-particle dispersion.
 We start with the case of circular electron pockets, and then extend the analysis to elliptical pockets.

\subsection{Circular electron pockets}

For circular pockets, $\xi_{1,\vec{k}} = \xi_{2,\vec{k}} = \xi_{k} = v_F (k-k_F)$  is independent on the angle along the FS.
The quadratic part of the Hamiltonian is diagonalized by the transformation to new operators $a$ and $b$  via $f_{1,2} = \!( a \mp b )\!/ \sqrt{2} $ and becomes
 \be
 H_2 = \sum_{\vec{k}} \xi^+_{{\vec{k}}} a_{\vec{k}}^{\dag} a_{\vec{k}} + \sum_{\vec{k}} \xi^-_{{\vec{k}}} b_{\vec{k}}^{\dag} b_{\vec{k}}
  \label{ch_5}
  \ee
  with $\xi^{\pm}_{{\vec{k}}} = \xi_{k} \pm \lambda$.
For angle-independent $\lambda$, the  new FSs remain concentric circles with different radii.

The BCS Hamiltonian takes the form
 \begin{align}\label{Hi_1}
H_{\mathrm{BCS}}  = & \Delta \sum_{\vec{k}}
\big\{  (a_{\vec{k}} a_{-{\vec{k}}} + b_{\vec{k}} b_{-{\vec{k}}} )
 \notag \\
 & +
(a_{\vec{k}} b_{-{\vec{k}}} + a_{\vec{k}} b_{-{\vec{k}}} ) \alpha \cos 2 \theta_{\vec{k}}
 \big\}  + h.c.
\end{align}
We see that, in terms of $a$ and $b$ operators, the $\cos 2 \theta$ term measures the strength of inter-band pairing.
As $\lambda$ increases and the two hybridized FSs become more separated, inter-pocket pairing becomes less important, and one expects that the angular dependence of the pairing interaction will play a lesser role.
This is what calculations show, as we demonstrate below.
\begin{figure}[h]
\begin{center}
\includegraphics[width=0.86\columnwidth]{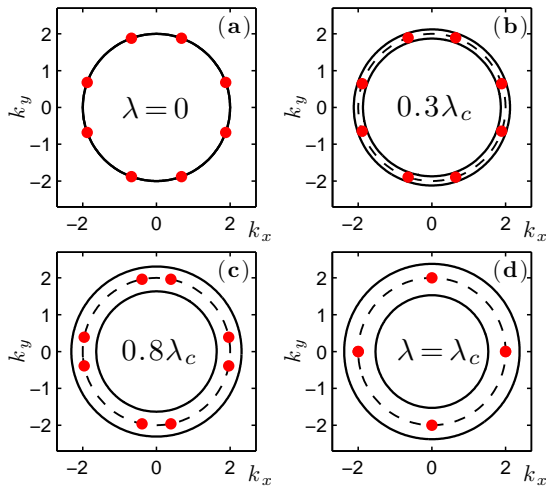}
\caption{Color online.
The evolution of the FS and gap nodes shown by (red) dots for circular pockets and constant (angle independent) hybridization amplitude. Without hybridization, the two electron FSs are identical, and each has four nodes (panel a). Once the hybridization parameter $\lambda$ becomes non-zero, the two electron FSs split, but the nodes remain on the original (non-hybridized) FS (panels b and c).  As $\lambda$ increases, the pairs of neighboring nodes come closer to each other and eventually merge at $\lambda_c$ (panel d) and disappear at larger
  $\lambda$.
  We set $\alpha = 1.3$. }
\label{fig:1}
\end{center}
\end{figure}

The quasi-particle dispersion is obtained from zeros of the inverse propagator
$G_{\omega,\vec{k}}^{-1} =  \omega I - M $
with  matrix $M$ and
 a unit
matrix $I$ operating in  Nambu space
 $\psi_{\vec{k}} = [ a_{\vec{k}}, a_{-\vec{k}}^{\dag}, b_{\vec{k}},  b_{-\vec{k}}^{\dag} ]^{tr}$.
We have
\begin{align}\label{M}
M = \begin{bmatrix}
\xi_+  &  \Delta & 0 & \Delta  y  \\
\Delta & -\xi_+   & \Delta  y &  0 \\
0 &  \Delta  y  & \xi_-   & \Delta \\
\Delta  y & 0 & \Delta & - \xi_-
\end{bmatrix} ,
\end{align}
where
 $y = \alpha \cos 2 \theta_{\vec{k}} $.
The nodal points are located at $\vec{k}_{n}$ for which $\det M(\vec{k}) =0$.
Equation (\ref{M}) gives
\begin{align}\label{detM}
\det M =
4 \Delta^2  \xi_{k}^2
+
\left[ \xi_{k}^2 - \lambda^2  + \Delta^2 ( y^2  - 1 ) \right]^2 \, .
\end{align}
\begin{figure}[h]
\begin{center}
\includegraphics[width=\columnwidth]{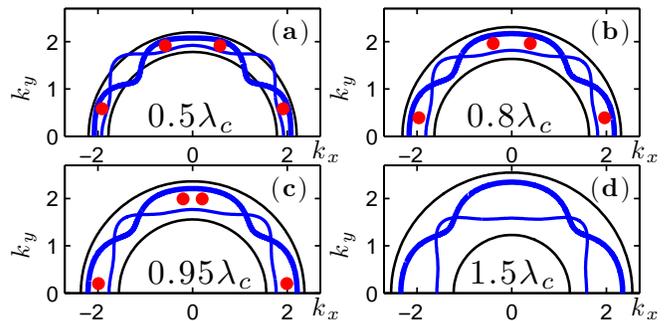}
\caption{Color online.
The location of the NSL, defined as locus of points for which fermion does not change its energy between the normal and the superconducting states.
 Thick (blue) lines are NSL which are detectable by ARPES (the peaks above and below $T_c$ are located at the same negative energy),  thin (blue) lines are NSL for which the peak above $T_c$ is at a positive energy.  The hybridization parameter $\lambda$ increases from (a) to (d). Dots (red) are nodal points at which quasiparticle energy is zero in a superconductor (but not in the normal state).
The initial evolution of NSL is quite involved. We consider it in Appendix B.}
 \label{fig:2}
\end{center}
\end{figure}
We see that the nodes reside on the ``bare'' FS, $\xi_{k}=0$, and their angular  position ${\theta}_{\vec{k}}$ is set by
\be
\label{theta_n}
\cos \theta_{\vec{k}} = \pm \frac{\sqrt{\lambda^2 + \Delta^2 }}{\alpha \Delta}\, .
\ee
 Equation~\eqref{theta_n} has eight solutions along a circle
 (4 pairs of nodes near  $\theta =0$ and other symmetry-related points
 $\theta = \pi/2, \theta =\pi$, and $\theta = 3\pi/2$).
As $\lambda$ increases, the two nodes located, e.g.,  near $\theta =0$ move toward each other, and at a critical $\lambda_c$ given by
\be
 \lambda_c = \Delta \sqrt{ \alpha^2 - 1} \, .
 \label{ch_3}
 \ee
the  neighboring nodes   merge along symmetry lines.  At larger $\lambda$ the nodes disappear.
 We show this behavior in Fig.~\ref{fig:1}. Observe that the critical $\lambda_c$ is of order $\Delta$, i.e., is rather small. We will see below that $\lambda_c$ is much larger when electron FSs are not circular.

In the presence of impurities, the nodes disappear already at $\lambda$ smaller than $\lambda_c$,
once the distance between the nodes  becomes smaller than some minimum value set by impurity scattering~\cite{peter,peter_1}.

Consider the range $\lambda < \lambda_c$ in more detail.
That the nodes are not located on the actual (hybridized) FSs has a profound
 effect on the ARPES spectrum.
In the normal state, the quasiparticle energies are $\xi_k \pm \lambda$, i.e., one energy is  positive along  $\xi_k =0$ and the other is negative.
Because  ARPES intensity is proportional to the Fermi function, ARPES will only detect a negative mode at $\omega = -\lambda$.
 That the nodes in the superconducting state are located along $\xi_{k} =0$ then implies that around these ${\vec k}$, the position of the maximum in the ARPES spectra shifts towards a {\it smaller} frequency as the temperature drops below $T_c$, instead of staying intact, as at a nodal point of a ``conventional'' superconductor with gap nodes.
At the same time, one can easily make sure that near the larger FS ARPES peak shifts to a larger negative frequency below $T_c$.
As a result, there exist  lines in k-space along which ARPES intensity does not shift upon cooling through $T_c$ (in the introduction we termed  these lines as "no-shift" lines, or NSL).

In the absence of hybridization, the NSL are radial beams along $y = \alpha \cos 2 \theta_k = \pm 1$ for which superconducting dispersion $\omega ({\vec k}) = -(\xi^2 + \Delta^2 (1 \pm y)^2)^{1/2}$ coincides with the normal state dispersion $-\xi$.
Once $\lambda$ becomes non-zero, $\omega ({\vec k})$ splits into two branches,
\begin{align}
\omega_{1,2}^2  =  ( \Delta^2 + y^2 \Delta^2 + \xi^2 + \lambda^2)
\mp 2 \sqrt{ S }   \, ,
\label{ch_1}
\end{align}
where $S = \xi^2 \lambda^2 +  y^2 \Delta^2 (\Delta^2 + \lambda^2) $. We analyzed (\ref{ch_1}) and found that the NSL rapidly undergo a series of bifurcations (see Appendix B) and for $\lambda \lesssim \lambda_c$
evolve into banana-like  loops located in between the two hybridized FSs and surrounding the actual nodal points (see Fig.~\ref{fig:2}).
The NSL persist even above $\lambda_c$, when the actual nodes disappear, and  should be easily detectable by ARPES (at least, in theory).
\begin{figure}[h]
\begin{center}
\includegraphics[width=\columnwidth]{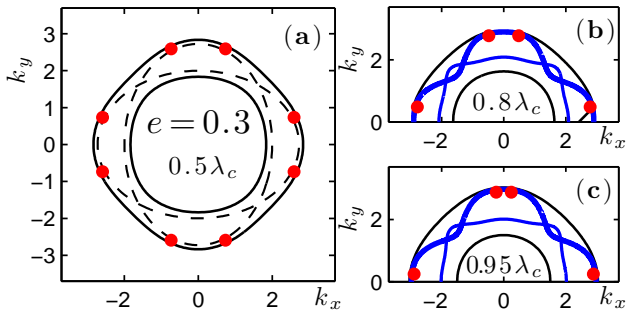}
\caption{
 Color online.
Hybridization of 2D elliptical FSs.
Panel (a)-- the FSs and the location of nodes (red dots).
Solid  and dashed black lines are the actual, hybridized FSs, and the original FSs, respectively.
Similarly to the case of
circular FSs, the pairs of nodal points come closer to each other as $\lambda$ increases, and merge and disappear at a critical $\lambda_c$.
Panels (b) and (c) -- the evolution of the
NSL
 with increasing $\lambda$.
Solid black lines show the hybridized FSs.
Thick and thin solid blue lines are NSL detectable and undetectable by ARPES, respectively.
 The 2D analysis is applicable to 1111 materials with simple tetragonal lattice structure. For 122 materials
 with body-cetered lattice structure, the situation is somewhat different, see Fig.~\protect\ref{fig:4}.}
\label{fig:3}
\end{center}
\end{figure}

\subsection{Non-circular Fermi pockets}

For non-circular electron pockets the quadratic Hamiltonian
 \eqref{H01}
is diagonalized by the angle-dependent transformation
$f_1 = u a + v b $, $f_2 = -v a + u b $, with
$(u,v) = (\cos \phi,- \sin \phi)$ and
$\cos 2 \phi = (\xi_1 - \xi_2)/\sqrt{ (\xi_1 - \xi_2)^2 + 4 \lambda^2 } $, $\sin 2 \phi =  2 \lambda / \sqrt{ (\xi_1 - \xi_2)^2 + 4 \lambda^2 }$.
 The diagonalization  yields
 \begin{align}\label{H22}
H_2 = \sum_k \xi_a a_k^{\dag} a_k +  \sum_k \xi_b b_k^{\dag} b_k \, ,
\end{align}
with
\begin{align}\label{xi_ab}
\xi_{a,b} =
\frac{ 1 }{ 2 } \left( \xi_1 + \xi_2 \right) \pm
\left[ \lambda^2 +  \left( \xi_1 - \xi_2 \right)^2/4 \right]^{1/2} \, .
\end{align}
The Nambu matrix $M$ becomes, instead of Eq.~\eqref{M},
\begin{align}\label{MF1}
\begin{bmatrix}
\xi_a     &  \Delta( 1 + \alpha_2) & 0 & \Delta \alpha_1  \\
\Delta( 1 + \alpha_2) & -\xi_a   & \Delta  \alpha_1 &  0 \\
0 &  \Delta   \alpha_1  & \xi_b    & \Delta( 1 - \alpha_2) \\
\Delta  \alpha_1 & 0 & \Delta( 1 - \alpha_2) & - \xi_b
\end{bmatrix} ,
\end{align}
where
$\alpha_1 = y \sin 2 \phi$, $\alpha_2 =y \cos 2 \phi$.
The location of the nodes is again specified by
$ \det M =0$. We have
\begin{align}
\det M = & \
 \Delta^2 \left[ ( \xi_a + \xi_b) + \alpha_2 (\xi_b - \xi_a) \right]^2
\notag \\
& +
\left[ \xi_a \xi_b + \Delta^2 ( y^2 - 1 ) \right]^2 \, .
\label{ch_2}
\end{align}
Both terms in (\ref{ch_2}) are non-negative and must vanish simultaneously
at the locations of the nodes.
This sets two conditions,
one on $\xi_{a,b}$ and one on the angle $\theta_k$.
Solving the coupled set we find the same behavior as for circular pockets -- the eight nodes are located in between the hybridized electron FSs, and the pairs of neighboring nodes move towards each other as $\lambda$ increases, merge along symmetry directions at a critical $\lambda_c$ and disappear at larger $\lambda$ (Fig.~\ref{fig:3}).
The critical $\lambda_c$ is given by
\be\label{crit1}
\lambda_c =  (\alpha^2 - 1)^{1/2} \left[ ( ({\bar \xi}_1 - {\bar \xi}_2 )/2)^2   + \Delta^2 \right]^{1/2}\, ,
\ee
where ${\bar \xi}_1$ and ${\bar \xi}_2$ are dispersions $\xi_{1,k}$, $\xi_{2,k}$ along one of four symmetric directions $\theta_k = 0,\pi/2,\pi,$ or $3\pi/2$, and the value of $k$ is fixed by the condition ${\bar \xi}_1 ={\bar \xi}_2 (\alpha -1)/(\alpha +1)$.
For elliptical pockets, with
$ \xi_{1,2} =  k_x^2 / 2 m_{1,2}  +  k_y^2 /2 m_{2,1 }-\mu$,
$({\bar \xi}_1 - {\bar \xi}_2 )/2 = \mu e/(1 - \alpha |e|)$, where $e= (m_1-m_2)/(m_1 + m_2)$ is the eccentricity.
 If the two FSs were nearly circular but of different radii before the hybridization (as in body-centered tetragonal 122 systems), i.e., $\xi_{1,k} = k^2/(2m_1) -\mu$, $\xi_{2,k} = k^2/(2m_2) -\mu$, and $m_1 = m(1 +\epsilon), ~m_2 = m(1-\epsilon)$,
 then ${\bar \xi}_1 - {\bar \xi}_2 =
 - k^2_F \epsilon /(m^2)$,
where $k_F \approx \sqrt{2 m \mu}$.
 In both cases, $\lambda_c$ obviously increases when electron pockets becomes non-identical.

We computed the dispersion by solving $\det (\omega (k) - M) =0$ and again found
that NSL form banana-shape closed loops around the true nodes.  We show the results in Fig.~\ref{fig:3}.
In similarity to the case of circular pockets, NSL survive even when the actual nodes disappear.
 \begin{figure}[h]
\begin{center}
\includegraphics[width=\columnwidth]{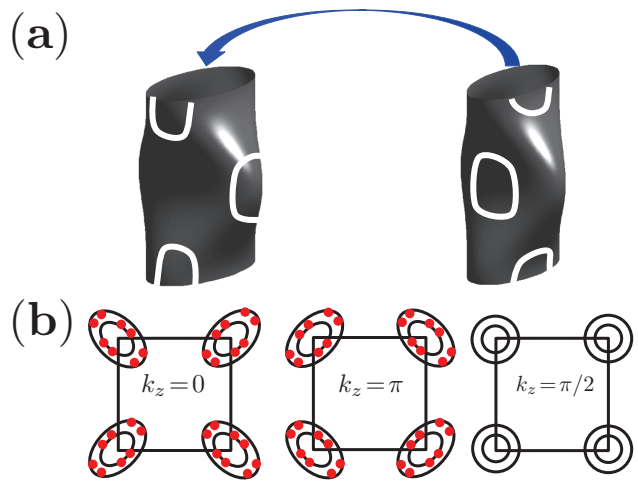}
\caption{Color online.
Panel (a) -- the 3D FSs of 122 systems, like BaFe$_2$(As$_{1-x}$P$_x$)$_2$, and the location of vertical loop nodes shown by white lines.
Two warped cylinders are shown separately, but the smaller one is actually inside the larger one, as the arrow indicates. Panel (b) --  cross section of the actual FSs at various $k_z$.
Red dots mark the location of the nodes.}
\label{fig:4}
\end{center}
\end{figure}

\section{The effect of hybridization on the nodes, 3D body-centered tetragonal lattice}
\label{sec:3D}

In 122 systems with body-center tetragonal lattice,  two electron pockets
  coupled by $\lambda$  are separated by $(\pi,\pi,\pi)$, where the third component is along $k_z$ direction~\cite{3D}.
  The
  pairs of
  pockets at $(0,\pi,0)$ and $(\pi,0,\pi)$, and at $(\pi,0,0)$ and
  $(0,\pi,\pi)$ are co-axial ellipses,
   which are well separated even without hybridization~\cite{mazin}.
The difference ${\bar \xi}_1 - {\bar \xi}_2$ is then large, and for $k_z$ near $0$ and near $\pi$ the critical $\lambda_c$  as given by Eq.~(\ref{crit1}),
is large and generally comparable to Fermi energy.
On the other hand, at $k_z =\pi/2$, the two electron pockets coupled by $\lambda$ are identical, and can be well approximated by  circles.
For these pockets, the critical $\lambda_c$ is given by (\ref{ch_3}) and is quite small, of order $\Delta$.
For $\lambda_c (\pi/2) <\lambda < \lambda_c (0,\pi)$, the  the nodes near $k_z =\pi/2$ are eliminated, and nodal lines form vertical loops, which are centered at $k_z =0$ and $k_z =\pi$ and close before reaching $k_z = \pi/2$.
We show this gap structure in Fig.~\ref{fig:4}.

\section{Application to P-containing pnictides.}
\label{sec:exp}

The three Fe-pnictide materials
 with hole and electron pockets,
for which experimental data strongly suggest the presence of gap nodes,
are  LaFeOP, with $T_c \leq 5$K (Ref.~[\onlinecite{lafop}]), the family
 BaFe$_2$(As$_{1-x}$P$_{x}$$)_2$ with the highest $T_c$ around 30K (Ref.~[\onlinecite{matsuda_1}]), and  LiFeP with $T_c \leq 5$K (Ref.~[\onlinecite{lifep}]).
All three materials contain phosphors.
 $k_z$-integrated probes like penetration depth, thermal conductivity, specific heat, and NMR~[\onlinecite{thermal_p}]
 all show the behavior consistent with line nodes.
In particular, thermal conductivity $\kappa$ scales linearly with $T$ at low $T$ and displays $\sqrt{H}$ behavior in a magnetic field, and $\lambda (T) - \lambda (0)$ is also linear in $T$ down to very low $T$.

The results of ARPES study of the gap structure are controversial.
Laser ARPES study~\cite{laser_arpes} probed the gap near the three hole pockets and found all of them almost angle independent, at least for $k_z$ probed by laser ARPES.
These authors argued that the nodes must be on electron pockets.
Synchrotron ARPES data were, on the other hand, interpreted as evidence for a horizontal line node at some $k_z$ on one of hole FSs.
We argue in Appendix~\ref{sec:app_B} that another fitting procedure of the data from Ref.~[\onlinecite{recent_arpes}], which, we believe, is more appropriate,
is consistent with some $k_z$ dispersion of the gap but no nodes on hole FSs.
And the most recent synchrotron ARPES data~\cite{arpes_latest} show
no nodes on hole pockets and strong gap variation on
electron pockets.
It  appears that more work is needed to resolve  the structure of the gaps on hole and electron FSs in ARPES studies.  From theory viewpoint, a horizontal node
 on a hole FS is possible~\cite{3D}, but less justified than nodes on electron pockets, as the latter appear quite naturally due to competition between
 inter-pocket repulsion between hole and electron pockets, which favors $s^{+-}$ superconductivity, and intra-pocket repulsion, which is against any superconductivity.
The gap on electron pockets acquires $\cos 2 \theta$ variations to reduce the effect of intra-pocket repulsion and allow superconductivity to develop~\cite{cvv,maier}.
This reasoning is consistent with the argument~\cite{kuroki_1} that a replacement of As by P changes the hight of a pnictide with respect to Fe plane, which effectively reduces inter-pocket electron-hole interaction, forcing the gap to develop nodes on electron pockets to reduce the effect of intra-pocket repulsion.
We therefore believe that nodes more likely reside on electron pockets, as we suggest in our analysis.

The structure of the nodes on electron pockets has been discussed in the context of the analysis of thermal conductivity data in
 BaFe$_2$(As$_{1-x}$P$_{x}$$)_2$.
Measurements of the oscillations of thermal conductivity as a function of a direction of a magnetic field have been reported recently~\cite{matsuda}, and $\cos 4 \theta$ component of these oscillations has
 been interpreted using  the same form of the gap on electron pockets as in our study: $\Delta (k_z) = \Delta (1 \pm \alpha (k_z) \cos 2 \theta)$. The best fit to the data yields $\alpha(k_z) >1$ for $k_z$ near $0$ and $\pi$ and $\alpha (k_z) <1$ for $k_z$ near $\pi/2$.
This form of $\alpha (k_z)$ implies that nodes form vertical loops centered at $k_z=0$ and $k_z = \pi$.
This is precisely what we found in our calculations.
We therefore argue that our calculation  provides  microscopic explanation of the appearance of vertical loop nodes in
 BaFe$_2$(As$_{1-x}$P$_{x}$$)_2$.

\section{Conclusion}
\label{sec:concl}
To conclude, in this paper we considered how the originally nodal $s^{+-}$ gap changes if we include into the effective tight-binding model for Fe atoms an additional term with momentum transfer $(\pi,\pi,0)$ in 1111 systems and $(\pi,\pi,\pi)$ in 122 systems.
We show that such a term is generally present because the hopping between Fe orbitals is primarily an indirect hopping via pnictogen (chalcogen) orbitals, and  pnictogen (chalcogen) atoms are  located above and below the Fe plane in a checkerboard order.
In 122 systems this order flips between neighboring Fe planes along z-axis.
This additional  hopping term hybridizes the two electron pockets and
affects the gap structure.
We found that the pairs of neighboring nodes (points were the quasiparticle energy is zero below $T_c$) approach each other as hybridization increases and disappear once the hybridization parameter $\lambda$ exceeds a certain  threshold $\lambda_c$.
We argued that in 122 systems, like BaFe$_2$(As$_{1-x}$P$_x$)$_2$,  the threshold value depends on $k_z$ and is much smaller near $k_z =\pi/2$ than near $k_z =0$ and $\pi$.
In this situation, at intermediate $\lambda$, the gap nodes form vertical loops which are centered at $k_z =0$ and $\pi$ and close
 up
 before reaching $k_z = \pi/2$.

We also found that in $k_z$ cross-sections where the nodes are present, they are located away from the hybridized FSs.
As a result, at a nodal point, the  peak in the ARPES energy distribution curve  shifts, upon cooling through $T_c$, from a negative frequency to a smaller frequency.
We showed that the nodes are surrounded by the "no-shift" lines -- the subset of $k-$points at which ARPES peak does not shift between $T > T_c$ and $T < T_c$.
These lines initially form beams along the directions where $s^{+-}$ gap vanishes, but they rapidly evolve as $\lambda$ increases, and for $\lambda \lesssim \lambda_c$ form banana-shape loops in $(k_x,k_y)$ plane around the nodes.
We propose to search for these  "no-shift" lines in ARPES measurements.

 We acknowledge useful conversations with D. Basov, L. Bascones,  S. Borisenko, A. Coldea, D. L. Feng, R. Fernandes,
 P. Hirschfeld, I. Eremin, A. Kordyuk,  S. Maiti, Y. Matsuda, I. Mazin, T. Shibauchi,
   R. Thomale, M. Vavilov, I. Vekhter, A. Vorontsov, and H.H. Wen.
We particularly thank I. Mazin for the discussion on the angular dependence of the hybridization amplitude.
This work was supported by the University of Iowa (M.K.) and by the Department of Energy grant DE-FG02-ER46900 (A.C.)

\begin{appendix}

\section{Microscopic mechanism of hybridization}
\label{sec:hyb}

In this Appendix  we present microscopic derivation of Eq.~\eqref{hyb} and obtain explicit expressions for $\lambda (\vec{k})$ in terms of microscopic parameters.
 We assume, following earlier works, that the hopping between Fe atoms occurs via pnictogen/chalcogen.
 Such a process gives rise to two types of Fe-Fe hopping terms: the ones with zero momentum transfer and the ones with momentum transfer $\vec{Q}$.
The terms with zero momentum transfer (and the ones with  momentum transfer $\vec{Q}$ between fermion states near the center and the corners of the 1FeBZ~\cite{hu_hao}) give rise to the electronic structure with hole and electron pockets.  We assume that these terms are already incorporated into the tight-binding Hamiltonian of Eq.~\eqref{H0},
and focus only on the terms with momentum transfer $\vec{Q}$, which involve fermions with momenta near  $(0,\pi)$ or $(\pi,0)$ and give rise to Eq.~(\ref{hyb}).

\begin{figure}[ht]
\begin{center}
\includegraphics[width=\columnwidth]{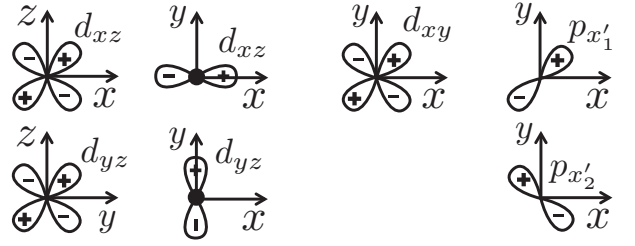}
\caption{Schematic representation of the three Fe orbitals $d_{xz}$, $d_{yz}$, $d_{xy}$ and $p_x$ and $p_y$ orbitals of As/Se.
 }
\label{fig:orbitals}
\end{center}
\end{figure}

Equation \eqref{hyb} implies that
\be\label{lambda:real}
\lambda_{\vec{k}} = \langle \psi_{1,\vec{k}} | H | \psi_{2,\vec{k} + \vec{Q}} \rangle\, ,
\ee
where $H$ is the full hopping Hamiltonian.
The band operators $\psi_{1,\vec{k}}$ and $\psi_{2,\vec{k}+\vec{Q}}$ are
linear combinations of atomic orbital operators.
We consider $d_{xz}$, $d_{yz}$ and $d_{xy}$ orbitals, and neglect
$d_{z^2}$ and $d_{x^2-y^2}$ orbitals which do not contribute to the states near the Fermi level~\cite{cao}.
Let $|f_{s,\vec{n}} \rangle$ denote orbital operators localized at site $\vec{n}$ with $s=1,2,3$ standing for $d_{xz}$, $d_{yz}$ and $d_{xy}$ respectively.
The corresponding wave functions are shown in  Fig.~\ref{fig:orbitals}.
The band operators are
 \begin{align}\label{real1}
| \psi_{1,\vec{k}} \rangle &= \sum_{s=1}^3  \gamma_{1}^{s} (\theta_{\vec{k}})   |f_{s,\vec{k}}\rangle\,
\notag \\
| \psi_{2,\vec{k}+\vec{Q}} \rangle &=  \sum_{s=1}^3  \gamma_{2}^{s} (\theta_{\vec{k}+ \vec{Q}})   |f_{s,\vec{k}+\vec{Q}}\rangle\, ,
\end{align}
where
\begin{align}\label{real1a}
|f_{s,\vec{k}}\rangle =
\frac{ 1 }{ \sqrt{N} } \sum_{\vec{n}} e^{ i \vec{k}\vec{n}}  |f_{s,\vec{n}}\rangle\, .
\end{align}
The $\vec{k}$-dependent coefficients $\gamma^s_{1}(\theta_{\vec{k}})$ ($\gamma^s_{2}(\theta_{\vec{k}})$)
specify orbital contents of the pockets centered at $(0,\pi)$ or $(\pi,0)$, respectively, and are input parameters for our consideration.

Consider momentarily a single layer.
We have  $\vec{n} = (n,m)$ and $\vec{Q} = (\pi,\pi)$.
Using
\be\label{Q}
e^{ i (\vec{k}+\vec{Q})\vec{n}}  = (-1)^{n+m} e^{ i \vec{k}\vec{n}}\, .
\ee
we can separate the sums over $\vec{n}$ into contributions from even and odd sublattices, i.e., split $\sum_{\vec{n}}$ into  $\sum'_{\vec{n}} + \sum''_{\vec{n}}$,
where the first (second) sum is limited to even (odd) values of $n+m$.
 Equation \eqref{real1} then takes the form
\begin{align}\label{real2}
| \psi_{1,\vec{k}} \rangle &= \frac{ 1 }{ \sqrt{N} }
\left[ \sum'_{\vec{n}}  + \sum''_{\vec{n}} \right]
e^{ i \vec{k}\vec{n}}  \sum_{s=1}^3  \gamma_{1}^{s} (\theta_{\vec{k}})   |f_{s,\vec{n}}\rangle\, ,
\notag \\
| \psi_{2,\vec{k}+\vec{Q}} \rangle &= \frac{ 1 }{ \sqrt{N} }
\left[ \sum'_{\vec{n}}  - \sum''_{\vec{n}} \right]
e^{ i \vec{k}\vec{n} } \sum_{s=1}^3  \gamma_{2}^{s} (\theta_{\vec{k}})   |f_{s,\vec{n}}\rangle\, .
\end{align}
Substituting \eqref{real2} into \eqref{lambda:real} we obtain
\begin{align}\label{real:3}
\lambda_{\vec{k}} = &
\frac{ 1 }{ N} \sum_{\vec{n}'}
\left[ \sum'_{\vec{n}} - \sum''_{\vec{n}} \right]
e^{i\vec{k}(\vec{n} - \vec{n}')}
\notag \\
\times &
\sum_{s,s'}\left[ \gamma_1^s(\theta_{\vec{k}}) \right]^*
 \gamma_{2}^{s'}(\theta_{\vec{k}})
\left\langle f_{s',\vec{n}'} \vert H \vert f_{s,\vec{n}} \right\rangle
\end{align}
The sites within each  sublattice are identical, hence $\left\langle f_{s',\vec{n}'} \vert H \vert f_{s,\vec{n}} \right\rangle$ depends only on ${\vec n}' - {\vec n}$.
Introducing
\begin{align}\label{AA'}
A^{\vec{l}}_{s's} &= \left\langle f_{s',\vec{n}+\vec{l}} \vert H \vert f_{s,\vec{n}} \right\rangle \quad n+m = 2p
\notag \\
A'^{\vec{l}}_{s's} &= \left\langle f_{s',\vec{n}+\vec{l}} \vert H \vert f_{s,\vec{n}} \right\rangle \quad n+m = 2p+1\, ,
\end{align}
where $p$ is an integer, we re-write \eqref{real:3} as
\begin{align}\label{real:4}
\lambda_{\vec{k}} = &
\frac{ 1 }{ 2} \sum_{\vec{l};s',s}
\left(A^{\vec{l}}_{s's} - A'^{\vec{l}}_{s's} \right)e^{- i\vec{k}\vec{l}}
\left[ \gamma_1^s(\theta_{\vec{k}}) \right]^*
 \gamma_{2}^{s'}(\theta_{\vec{k}})\, .
\end{align}
The summation in Eq.~\eqref{real:4}
formally runs over all ${\vec l}$ but in reality
does not extend beyond second neighbors. The same expression is obtained in the 3D case, but then the sum over ${\vec l}$ extends to neighbors in XY plane and along $z$ axis.

Equation~\eqref{real:4} expresses $\lambda_{\vec{k}}$ in terms of the band structure
 parameters and hopping amplitudes.
We see that the hybridization parameter is non-zero only if
the hopping amplitudes are different for even and odd sublattices.

Further analysis requires the evaluation of the orbital hopping amplitudes and  the knowledge of the orbital content of band operators.
As we said, we neglect direct hopping between iron atoms and focus on a second order tunneling processes via pnictogene/chalcogene.
The amplitudes \eqref{AA'} are then expressed in terms of hopping integrals from a $d-$orbital on a Fe site to one of three $p-$ orbitals on a neighboring As/Se site.
The $p_z$ orbital has much smaller overlap integral compared to the two in-plane orbitals,  and we neglect it.
The in-plane  orbitals are presented in Fig.~\ref{fig:orbitals}.
Their wave functions are maximized in the direction along
$x'_1 = (x+y)/ \sqrt{2}$ and $x'_{2} = (-x+y)/ \sqrt{2}$, and we denote then by  $p_{x'_1,x_2'}$ and $\bar{p}_{x'_1,x_2'}$
for pnictogen/chlacogen  above and below the iron plane, respectively.

\begin{figure}[h]
\begin{center}
\includegraphics[width=1.0\columnwidth]{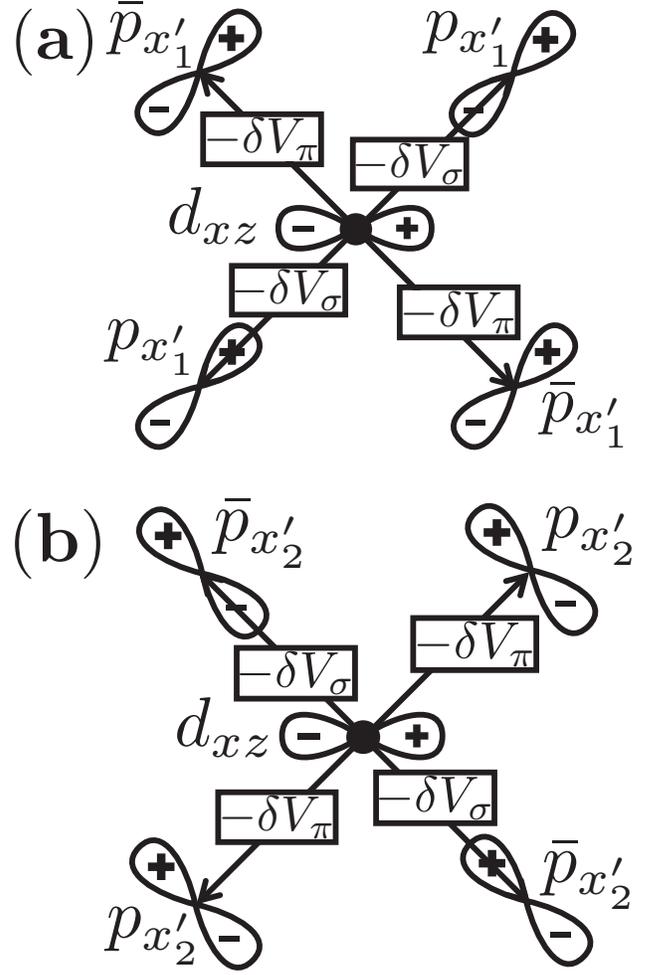}
\caption{
The overlap integrals between iron orbital $d_{xz}$, and in-plane
As/Se $p$ orbitals are defined by two parameters $\delta V_{\pi}$ and $\delta V_{\sigma}$.
We show Fe atom from even sublattice, for which 
 As/Se atoms  are displaced off the Fe plane by $+\delta$ 
  along the main diagonal ($x=y$),
and  by $- \delta$ along $x=-y$.  
The p-orbitals at $+\delta$ are labeled as $p$ and the ones at $-\delta$ as $\bar{p}$.
We assume that $\delta \ll 1$, in which case the overlap integrals are linear in $\delta$. 
(a) overlap between $d_{xz}$ and $p_{x'_1}$ orbitals,
(b) overlap between $d_{xz}$ and $p_{x'_2}$ orbitals.
}
\label{fig:overlaps_dxz}
\end{center}
\end{figure}

\begin{figure}[h]
\begin{center}
\includegraphics[width=1.0\columnwidth]{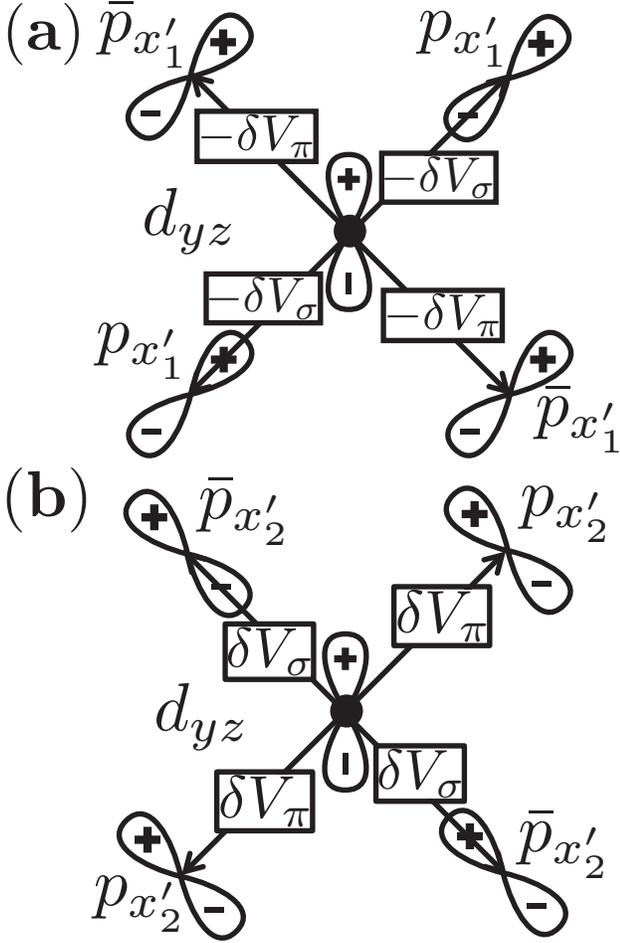}
\caption{
Same as in Fig. \ref{fig:overlaps_dxz} but for $d_{yz}$ orbital of Fe.
(a) overlap between $d_{yz}$ and $p_{x'_1}$ orbitals,
(b) overlap between $d_{yz}$ and $p_{x'_2}$ orbitals.
}
\label{fig:overlaps_dyz}
\end{center}
\end{figure}

Within the tight-binding approximation the hopping parameters between Fe and As/Se sites can be specified by a set of overlap integrals
$\delta V_\sigma$ and $\delta V_\pi$, as shown
in Figs.~\ref{fig:overlaps_dxz} and \ref{fig:overlaps_dyz}.
Here $\delta$ is the the deviation of As/Se atom an Fe plane.
The wave functions of $d_{xz}$ and $d_{yz}$ orbitals change sign under $\delta \to -\delta$, hence the overlap integrals with $p_{x_i}$ and ${\bar p}_{x_i}$ have
 opposite signs.

In what follows we consider first the model with only $d_{xz}$ and $d_{yz}$ orbitals.
We show that the hybridization parameter vanishes in a 2D case and in  3D systems for 1111 lattice structure, but is generally non-zero for 3D systems with 122 lattice structure, except special directions in ${\vec k}$ space where $\lambda_{\vec k}$ vanishes.
We then include into consideration $d_{xy}$ orbital (i.e., consider three-orbital model) and show that in this case the hybridization parameter  is
non-zero already in 2D and in 1111 systems.
For 122 systems, $\lambda_{\vec k}$ in the three-orbital model is less anisotropic than in the two-orbital model and  non-zero along all directions of ${\vec k}$.

\subsection{Two-orbital model}

We follow earlier work \cite{raghu} and assume that the pocket at $(0,\pi)$ is predominantly made out of $d_{xz}$ orbital, and the one at $(\pi,0)$ is made out of $d_{yz}$ orbital, i.e., in our notations
$\gamma_1^s (\theta_{\vec{k}}) = \delta_{s,1}$ and $\gamma_2^s (\theta_{\vec{k}}) = \delta_{s,2}$.

\subsubsection{Single Fe layer}

\begin{figure}[ht]
\begin{center}
\includegraphics[width=1.0\columnwidth]{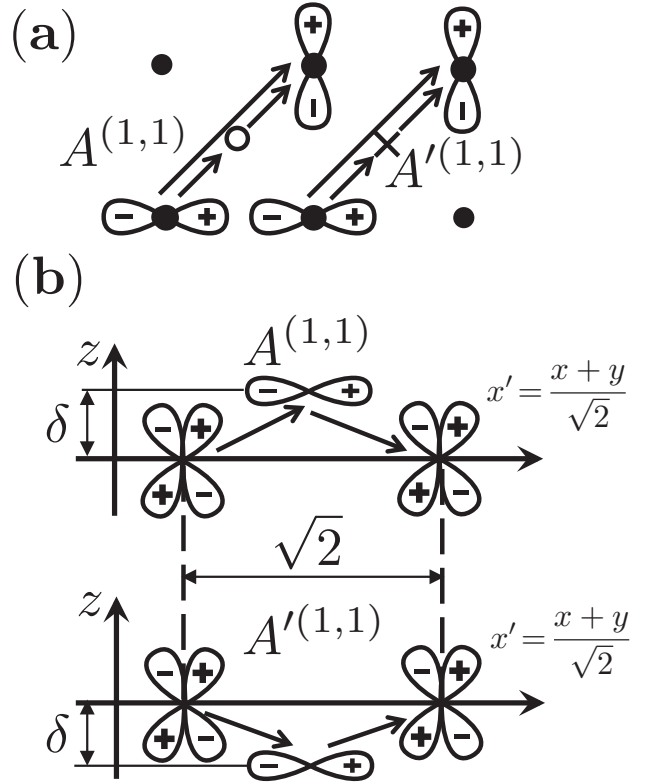}
\caption{The second order hopping processes within a single iron layer shown as a top view (a) and side view (b).
The equality of the two amplitudes, $A^{(1,1)}=A'^{(1,1)}$ follows from the observation that both orbitals, $d_{xz}$ and $d_{yz}$ are odd under the reflection
$z \rightarrow - z$.}
\label{fig:two}
\end{center}
\end{figure}

For a single iron layer, Eq.  \eqref{real:4}  simplifies to
\begin{align}\label{real:5}
\lambda_{\vec{k}} = &
\frac{ 1 }{ 2}
 \sum_{\vec{l}}
\left(A^{\vec{l}}_{2,1} - A'^{\vec{l}}_{2,1} \right)e^{- i\vec{k}\vec{l}} \, .
\end{align}
where, we remind, $A^l_{2,1}$ and  $A'^{\vec{l}}_{2,1}$ are hopping amplitudes from $d_{xz}$ orbital at site $\vec{n}$ to $d_{yz}$ orbital at cite $\vec{n} +\vec{l}$ staring from $\vec{n}$ at even or odd sublattice, respectively.
Each amplitude describes a two-stage process: the electron first hops from  $d_{xz}$ orbital to one of the two $p$ orbitals of pnictogen/chalcogen, and then hops from this $p$ orbital to  $d_{yz}$ orbital on either first or second neighbor,
see Fig.~\ref{fig:two}(a).
Because both $d_{xz}$ and $d_{yz}$ orbitals are odd in $z$, elementary amplitudes of hopping from $d$ to $p$ have different sign
 for hopping originating from even and odd sites. However, taken to second order, the amplitudes $A^l_{2,1}$ and  $A'^{\vec{l}}_{2,1}$
  turn out to be completely equivalent. We illustrate the equivalence in Fig.~\ref{fig:two} for $\vec{l} = (1,1)$.
   As a consequence,
 $A^{\vec{l}}_{2,1} = A'^{\vec{l}}_{2,1}$, i.e., $\lambda_k$ vanishes.

For completeness, we also computed the hybridization amplitude between $d_{xz}$ and $d_{yz}$ orbitals for $\vec{Q}=0$: 
\begin{align}\label{H12}
H_{1,2} = \sum_{\vec{k}} h_{1,2} (\vec{k}) f_1^{\dag} (\vec{k}) f_2 (\vec{k})\, .
\end{align} 
This term has been computed before~\cite{cao} and is part of the bare Hamiltonian which gives rise to hole and electron pockets.
Using our method, we  reproduce earlier result for $h_{1,2}$, as we now demonstrate.

Performing the same computations that for ${\vec Q} = (\pi,\pi)$ lead to Eq. \ref{real:5}, we
obtain for $\vec{Q}=0$ 
\be\label{h12}
h_{1,2} =
\frac{ 1 }{ 2}
 \sum_{\vec{l}}
\left(A^{\vec{l}}_{2,1} + A'^{\vec{l}}_{2,1} \right)e^{- i\vec{k}\vec{l}} \, .
\ee
%
 Because $A^{\vec{l}}_{2,1} = A'^{\vec{l}}_{2,1}$,
  Eq.~\eqref{h12} becomes
\be\label{h12_1}
h_{1,2} =
 \sum_{\vec{l}}
A^{\vec{l}}_{2,1} e^{- i\vec{k}\vec{l}} \, .
\ee
In the second order perturbation theory we have for, e.g., ${\vec l} = (\pm 1, \pm 1)$  (see Fig. \ref{fig:overlaps_dxz}, \ref{fig:overlaps_dyz})
\begin{align}\label{Q_zero}
A^{(1,1)}_{2,1} & = 
 A^{(-1,-1)}_{2,1} = - A^{(1,-1)}_{2,1} = -A^{(-1,1)}_{2,1} 
 \notag \\
&= \frac{ \delta^2 } { E_g} ( V_{\sigma}^2 - V_{\pi}^2 )\, ,
\end{align}
where the energy denominator $E_g$ is the energy separation between 3d to 4p orbitals. 
Substituting Eq.~\eqref{Q_zero} into Eq.~\eqref{h12_1}
 we obtain
\be
h_{1,2} = - 4  \frac{ \delta^2 } { E_g} ( V_{\sigma}^2 - V_{\pi}^2 ) \sin k_x \sin k_y + \ldots\, ,
\ee
where dots stand for the contributions from other ${\vec l}$.  
This result is in full agreement with 
earlier calculations, see e.g., Refs. 
\onlinecite{cao,review,hu_hao}.

\subsubsection{3D crystals, 1111 materials}

For 3D systems  with $1111$ structure,  Fe layers at different $z$ are al equivalent, and 3D folding vector $\vec{Q} = ( \pi, \pi,0)$.
 The arguments displayed in previous section apply to this case as well, i.e., for two orbital model $\lambda_k =0$, even if we include into consideration inter-layer tunneling.

\subsubsection{3D crystals, 122 materials.}

In $122$ materials with body-centred tetragonal crystal structure the situation is  qualitatively different.
The first observation is that the hybridization vector is $\vec{Q}=(\pi,\pi,\pi)$ because even and odd sublattices are formed by Fe atoms located at  $\vec{n} =(n,m,p)$ with $n+m+p$ even or odd, respectively.
This can be also be understood by noticing that the Fe-only lattice is simple cubic, but, because of As/Se,
Fe lattice has an fcc structure with the basis, or alternatively a rock salt structure.
The  folding vector $\vec{Q} = (\pi,\pi,\pi)$ appears as additional Bragg peak due to the transition from a simple cubic to an fcc lattice.
Without interlayer tunneling, $\lambda_{\vec k}$ is still zero, but inter-layer tunneling makes it finite as we show below.

\begin{figure}[ht]
\begin{center}
\includegraphics[width=\columnwidth]{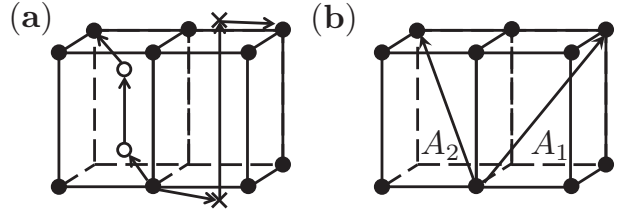}
\caption{Interlayer tunneling processes leading to a finite hybridization, \eqref{lambda:3D} in 122 material. }
\label{fig:3Dhybr}
\end{center}
\end{figure}

Inter-layer tunneling in real systems is a complex process which at least partly involves Ba atoms (Ref. \cite{mazin}).
We will avoid this complication and consider a toy model in which there is a direct tunneling between pnictogen/chalcogen atoms located at the same $(x,y)$ in different layers.
Because the position of pnictogen/chalcogen atoms relative to Fe plane oscillates along $z$ direction, there are two different tunneling amplitudes for such processes: the  one between pnictogen/chalcogen located above $n$-th plane and below $(n+1)$-th plane, and the other between pnictogen/chalcogen
located below $n$-th plane and above $(n+1)$-th plane (see Fig. \ref{fig:3Dhybr}).

Another peculiarity of 122 systems is that  the eccentricity of electron FSs in  122 systems changes sign  between $k_z =0$ and $k_z=\pi$. i.e., the FS at $(0, \pi 0)$ is elongated along the same direction as the FS at $(\pi,0, \pi)$.
In the full model, this is  due to the change of the relative weight of $d_{xy}$ orbital between $k_z=0$ and $k_z=\pi$ (Ref.~\cite{mazin}).
In two-orbital approximation,
we model the change of sign of eccentricity  by requiring that $d_{xz}$ and $d_{yz}$  orbitals interchange between $k_z =0$ and $k_z =\pi$ [i.e., we require that at $k_z=0$, the pocket at $(0,\pi)$ is made out of $d_{xz}$ orbital and is elongated along y-axis, and the one at $(\pi,0)$ is made out of $d_{yz}$ orbital and is elongated along $x-$axis, while at $k_z=\pi$, the pocket at $(0,\pi)$ is made out of $d_{yz}$ orbital and is elongated along $x$, and the one at $(\pi,0)$ is made out of $d_{xz}$ orbital and is elongated along $y$]. The hybridization term  with ${\vec Q} = (\pi,\pi,\pi)$  then connects $(0,\pi)$ pocket at $k_z=0$ and $(\pi,0)$ pocket at $k_z = \pi$, which have the same orbital character, i.e., $s$ and $s'$ in the Eq. \eqref{real:4}  for $\lambda_{\vec k}$ are the same.

The process which gives rise to a non-zero $\lambda_{\vec k}$ is a three-stage process in which an electron  from $d_{xz}$ or $d_{yz}$ orbital on an even sublattice in layer $n$ hops to one of two p-orbitals on a pnictogen/chalcogen, then hops vertically to a pnictogen/chalcogen located near
 next Fe layer, and then hops to the same $d-$orbital on that layer.
 One can easily verify that the terms in $\lambda_{\vec k}$ in which do not cancel out between even and odd sublattices are the ones with
  $\vec{l} = (\pm 1, \pm 1, \pm 1)$.
  There are two different
 hopping amplitudes for these momenta (see Fig. \ref{fig:3Dhybr}): for even sublattice we have
$A_{ss}^{(1,1,1)} = A_{ss}^{(1,-1,-1)} = A_{ss}^{(-1,-1,1)}=A_{ss}^{(-1,1,-1)} = A_1$,
 $A_{ss}^{(-1,1,1)} = A_{ss}^{(1,-1,1)} = A_{ss}^{(1,1,-1)}=A_{ss}^{(-1,-1,-1)} = A_2$, where $s=1,2$, and for odd sublattice
 $A_{ss}^{' (1,1,1)} = A_{ss}^{' (1,-1,-1)} = A_{ss}^{' (-1,-1,1)}=A_{ss}^{' (-1,1,-1)} = A_2$,
 $A_{ss}^{' (-1,1,1)} = A_{ss}^{' (1,-1,1)} = A_{ss}^{' (1,1,-1)}=A_{ss}^{' (-1,-1,-1)} = A_1$.
 Substituting these amplitudes into Eq.  \ref{real:4}, we see that in the real part of $\lambda_k$ the contributions from even and odd sublattices cancel out, but in the imaginary part of $\lambda_k$ they add up, such that
\be\label{lambda:3D}
\lambda_{\vec{k}} = - 4 i (A_1 - A_2 ) \sin k_x \sin k_y \sin k_z\, .
\ee
The expression \eqref{lambda:3D} is odd in all three momenta. This result is expected because the three reflection symmetries have been
 broken by hybridization.
The overall factor of $i$ reflects the fact that time reversal symmetry is not broken (i.e., $\lambda_k = \lambda^*_{-k}$).

For completeness, we also computed the hybridization term  between $d_{xz}$  and $d_{yz}$ orbitals and found it is also non-zero, but this time $\lambda_{\vec k}$ is real and is an even function of ${\vec k}$.

\subsection{Three orbital model}

We next consider how $\lambda_{\vec k}$ changes if we consider more realistic situation when  $d_{xy}$ orbital also contributes to the states near the FS.

\subsubsection{2D case}

We use as an input the results of previous calculations~\cite{review}, which found that the FS at $(0,\pi)$ is constructed out of $d_{xz}$ and $d_{xy}$
 orbitals, and  the one at $(\pi,0)$ is constructed out of $d_{yz}$ and $d_{xy}$ orbitals.  To a reasonable approximation,
 $\gamma^1_1 (\theta_k) = \sin \theta$,  $\gamma^3_1 (\theta_k) = \cos \theta$, and  $\gamma^i_2 (\theta) = \gamma^i_1 (\pi/2-\theta)$, where $\theta$ is counted from $y$-axis (as written, the formulas are valid in the first quadrant, for $0< \theta <\pi/2$).
 Accordingly
 \begin{align}\label{C1}
\psi_{1,\vec{k}}  &= f_{1,\vec{k}} \sin \theta   + f_{3,\vec{k}} \cos\theta
\notag \\
\tilde{\psi}_{2,\vec{k}+ {\vec Q}}  &= f_{2,\vec{k} + \vec{Q}} \cos\theta + f_{3,\vec{k} + {\vec Q}} \sin \theta
\end{align}
 The hybridization amplitude between $f_{1,\vec{k}}$ and $f_{2,\vec{k} + \vec{Q}}$ vanishes, as we found before, but we show in this section that
 the amplitudes between $f_{1,\vec{k}}$ and $f_{3,\vec{k} + {\vec Q}}$ and between $f_{3,\vec{k}}$ and $f_{2,\vec{k} + \vec{Q}}$ are non-zero.
In our notations, we then have
\begin{align}\label{3orb}
\lambda_{\vec{k}} = \lambda^{3,1}_{\vec{k}} + \lambda^{2,3}_{\vec{k}}\, ,
\end{align}
where
\begin{align}\label{lambda:13}
\lambda^{3,1}_{\vec{k}}  = \sin^2 \theta  \sum_{\vec{l}}
\left(A^{\vec{l}}_{3,1} - A'^{\vec{l}}_{3,1} \right)\frac{ e^{- i\vec{k}\vec{l}} }{ 2 }
\end{align}
and
\begin{align}\label{lambda:23}
 \lambda^{2,3}_{\vec{k}} =  \cos^2 \theta
 \sum_{\vec{l}} \left(A^{\vec{l}}_{2,3} - A'^{\vec{l}}_{2,3} \right)\frac{ e^{- i\vec{k}\vec{l}} }{ 2 }.
\end{align}

Because the wave function for  $d_{xy}$ orbital is even under the reflection $z \rightarrow - z$ an the ones  for
$d_{xz}$ and $d_{yz}$ orbitals are odd, $A'^{\vec{l}}_{3,1} = - A^{\vec{l}}_{3,1}$ and $A'^{\vec{l}}_{2,3} = - A^{\vec{l}}_{2,3}$.
Equation \eqref{lambda:23} then reduces to
\begin{align}\label{lambda:3orb}
\lambda_{\vec{k}} = \sum_{\vec{l}} ( \sin^2 \theta A^{\vec{l}}_{3,1} + \cos^2 \theta A^{\vec{l}}_{2,3})e^{- i \vec{k} \vec{l}}\, .
\end{align}
We now need to prove that this expression is non-zero.
%
\begin{figure}
\begin{center}
\includegraphics[width=1.0\columnwidth]{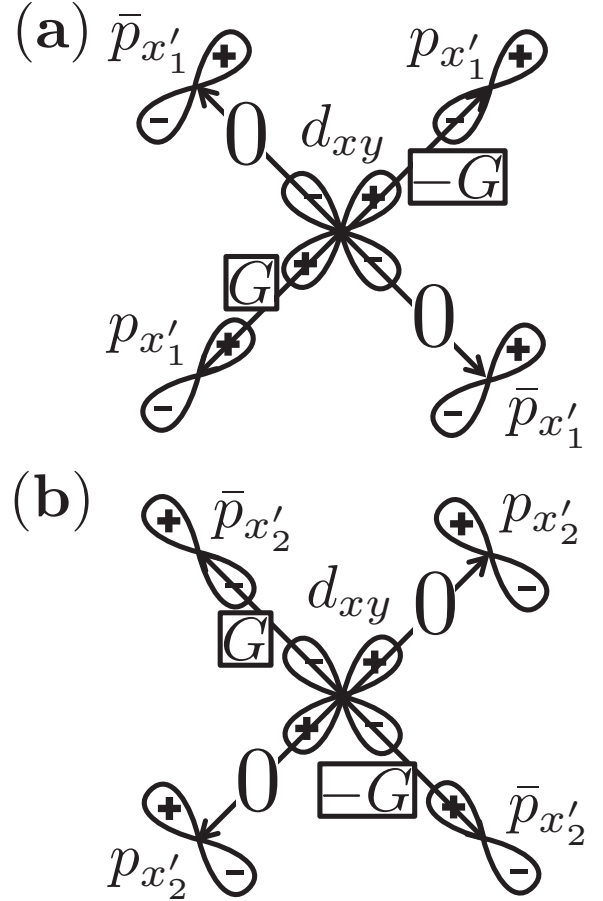}
\caption{Overlaps of iron $d_{xy}$ orbital with As/Se orbitals  $p_{x'_1}$  (a) and
$p_{x'_2}$  (b).
In the case (a) the tunneling is possible only along the main diagonal direction, $x=y$, while in the case (b) it occurs only along the perpendicular direction $x = - y $.
}
\label{fig:overlaps_xy}
\end{center}
\end{figure}
%
As before, we consider hopping between Fe sites as a two-stage process via neighboring p-orbitals.
 This process gives rise to hopping to nearest and next nearest neighbors on the iron lattice, i.e.,
  we need to consider ${\vec l} = (l_x, l_y)$ with $l_x, l_y = 0, \pm 1$. The overlap integrals with the $p$ orbitals on As/Se have been defined in
 Figs.~\ref{fig:overlaps_dxz} and \ref{fig:overlaps_dyz}.
In explicit form
\begin{align}\label{overlaps}
\left\langle p_{x'_1,\vec{n}+(1/2)(\hat{x} + \hat{y})}| H | f_{1,\vec{n}}\right\rangle & = -\delta V_{\sigma},
\notag \\
\left\langle  p_{x'_2,\vec{n}+(1/2)(\hat{x} + \hat{y}) }  | H |  f_{1,\vec{n}}\right\rangle & = -\delta V_{\pi}, \notag \\
\left\langle {\bar p}_{x'_1,\vec{n}+(1/2)(\hat{x} + \hat{y})}| H | f_{1,\vec{n}}\right\rangle & = \delta V_{\sigma},
\notag \\
\left\langle  {\bar p}_{x'_2,\vec{n}+(1/2)(\hat{x} + \hat{y}) }  | H |  f_{1,\vec{n}}\right\rangle & = \delta V_{\pi},
\end{align}
where, we remind, $\delta$ is  the deviation of As/Se from an Fe plane.

We define overlap integrals between $d_{xy}$ and $p$ orbitals  in Fig. ~\ref{fig:overlaps_xy} as
\begin{eqnarray}\label{overlaps1}
&&\left\langle  f_{3,\vec{n} + \hat{x}}   | H | p_{x'_2,\vec{n} +(1/2)(\hat{x} + \hat{y}) } \right\rangle = G,
\nonumber \\
&&\left\langle  f_{3,\vec{n} + \hat{x} }  | H |  p_{x'_1,\vec{n} +(1/2)(\hat{x} - \hat{y}) }  \right\rangle
 = G,\nonumber \\
&&\left\langle  f_{3,\vec{n} + \hat{x}}   | H | {\bar p}_{x'_2,\vec{n} +(1/2)(\hat{x} + \hat{y}) } \right\rangle = G,
\nonumber \\
&&\left\langle  f_{3,\vec{n} + \hat{x} }  | H |  {\bar p}_{x'_1,\vec{n} +(1/2)(\hat{x} - \hat{y}) }  \right\rangle = G\,
\end{eqnarray}
Other overlap integrals between $d_{xy}$ and in-plane $p-$orbitals vanish,

Using these notations,
 we obtain after a simple algebra,
\begin{align}\label{A_31_nn}
A_{3,1}^{(1,0)} = &
\langle f_{1,\vec{n}} | H | p_{x'_2,\vec{n} +(1/2)(\hat{x} + \hat{y}) } \rangle
\notag \\
& \times
\frac{1}{ E_g }
\langle   p_{x'_2,\vec{n} +(1/2)(\hat{x} + \hat{y}) } | H | f_{3,\vec{n} + \hat{x} }\rangle
\notag \\
 +&
\langle f_{1,\vec{n}} | H | p_{x'_1,\vec{n} +(1/2)(\hat{x} - \hat{y}) } \rangle
\notag \\
& \times
\frac{1}{ E_g }
\langle   p_{x'_1,\vec{n} +(1/2)(\hat{x} - \hat{y}) } | H | f_{3,\vec{n} + \hat{x} }\rangle \, .
\end{align}
\begin{align}
\label{Annn}
A_{3,1}^{(1,1)} & = \langle f_{1,\vec{n}} | H | p_{x'_1,\vec{n} +(1/2)(\hat{x} + \hat{y}) } \rangle
\notag \\
\times &
\frac{1}{ E_g }
\langle   p_{x'_1,\vec{n} +(1/2)(\hat{x} + \hat{y}) } | H | f_{3,\vec{n} + \hat{x}+\hat{y}} \rangle\, .
\end{align}
and similar results for other $A_{3,1}^{(l_x,l_y)}$  and $A_{2,3}^{(l_x,l_y)}$.
Substituting the expressions for the overlap integrals, Eqs. (\ref{overlaps}), (\ref{overlaps1}), we obtain
\begin{align}\label{A3}
A_{3,1}^{(\pm1,0)} = \mp \frac{2 \delta V_{\pi} G }{E_g} \, ,\,\,\,\,
A_{3,1}^{(0,\pm1)} = 0\, ,
\end{align}
\be\label{A4}
A_{3,1}^{(\pm1,\pm1)} = \mp \frac{ \delta V_{\sigma} G }{E_g} \, , \,\,\,\,
A_{3,1}^{(\pm1,\mp1)} = \mp \frac{ \delta V_{\sigma} G }{E_g}\, ,
\ee
and similarly
\begin{align}\label{A5}
A_{2,3}^{(0,\pm1)} = \mp \frac{2 \delta V_{\pi} G }{E_g} \, ,\,\,\,\,
A_{2,3}^{(\pm1,0)} = 0\, ,
\end{align}
\be\label{A6}
A_{2,3}^{(\pm1,\pm1)} = \pm \frac{ \delta V_{\sigma} G }{E_g} \, , \,\,\,\,
A_{2,3}^{(\pm1,\mp1)} = \mp \frac{ \delta V_{\sigma} G }{E_g}\, .
\ee
Substituting Eqs.~\eqref{A3}, \eqref{A4}, \eqref{A5} and \eqref{A6} in
Eq.~\eqref{lambda:3orb} and summing over four nearest and four next-nearest neighbors, we obtain
\begin{align}\label{3orb:res}
\lambda_{\vec{k}} &= \frac{ 4 i \delta G}{ E_g}
[V_{\pi} ( \sin^2 \theta \sin k_x + \cos^2 \theta \sin k_y )
\notag \\
+ &
V_{\sigma}( \sin^2 \theta \sin k_x \cos k_y - \cos^2 \theta \sin k_y \cos k_x )]\, .
\end{align}
Introducing small ${\tilde {\vec k}} ={\vec k} - (0,\pi)$, using the fact that for a small-size  pocket
$\cos \theta = {\tilde k}_y/|{\tilde k}|$, $\sin \theta = {\tilde k}_x/|{\tilde k}|$, and $\sin k_x \approx {\tilde k}_x$, $\sin k_y \approx - {\tilde k}_y$,
and extending the analysis to other quadrants (i,.e., to negative ${\tilde k}_x$ and ${\tilde k}_y$), we finally obtain
 \begin{align}\label{3orb:res_1}
\lambda_{\vec{k}} &= \frac{ 4 i \delta G}{ E_g |{\tilde k}|^2}
\left(V_{\pi} - V_\sigma\right) \left(|{\tilde k}|^3_x - |{\tilde k}|^3_y\right)\, .
\end{align}
We see that hybridization parameter $\lambda_{\vec{k}}$ is generally non-zero, except for diagonal directions ${\tilde k}_x = \pm {\tilde k}_y$.
This agrees with the result of numerical calculations~\cite{amalia}. The authors of Ref. \onlinecite{amalia} found that in the presence of spin-orbit interaction
$|\lambda|$ remains non-zero even along the diagonal directions.

\subsubsection{3D case}

In 1111 systems  the hybridization vector is $\vec{Q} = (\pi,\pi,0)$, and for
 weak inter-layer tunneling  $\lambda_{\vec k}$ weakly depends on $k_z$ and is nearly the same as in Eq. \ref{3orb:res}.

In 122 systems the hybridization vector is $\vec{Q} = (\pi,\pi,\pi)$, and the full expression for $\lambda_{\vec k}$ in 3D case is the sum
 of  $d_{xz}-d_{yz}$ contribution, Eq.~\eqref{lambda:3D}, and the contribution from the processes involving $d_{xy}$, Eq. \eqref{3orb:res}.
 The two-orbital contribution is
strongly $k_z$-dependent, while the three-orbital contribution is $k_z$-independent if we neglect inter-layer hopping, and weakly depends on $k_z$ if we include it.
Because the two contributions vanish along different symmetry directions, the total hybridization parameter is non-vanishing for all $\vec{k}$, though it has minima at $k_z =0,\pi$, and along $k_x =\pm k_y$ and $k_x =0, k_y =0$ in a given cross-section of $k_z$ ($k_x, k_y$ are measured with respect to a center of a pocket).
  The anisotropy of $\lambda_k$ is further reduced if we add spin-orbit interaction~\cite{mazin,amalia}.

\section{Evolution of the nodal lines at small hybridization}
\label{sec:app_A}

\begin{figure}[ht]
\begin{center}
\includegraphics[width=\columnwidth]{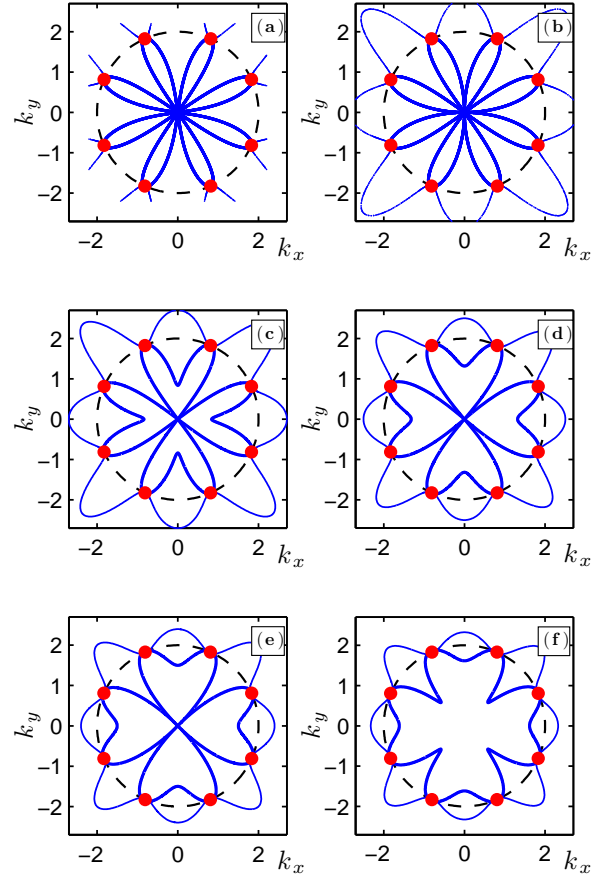}
\caption{Color online.
The initial evolution of the
 NSL defined as locus of points for which fermion does not change its energy between the normal and the superconducting states.
Dashed (black) lines show the FS before the hybridization.
 Thick (blue) lines are
 NSL
 which are detectable by ARPES (the peaks above and below $T_c$ are located at the same negative energy),  thin (blue) lines are ``nodal lines'' for which the peak above $T_c$ is at a positive energy.  The hybridization parameter $\lambda$ increases from (a) to (f). Dots (red) are nodal points at which quasiparticle energy is zero in a superconductor (but not in the normal state). Panel (a)  $\lambda < \lambda_{c,1}$, panel (b)  $\lambda = \lambda_{c,1}$, panels (c,d)  $\lambda_{c,1} < \lambda <\lambda_{c,2}$,  panel (e)  $\lambda = \lambda_{c,2}$, panel (f)
 $\lambda  >  \lambda_{c,2}$.
 Both $\lambda_{c,1}$ and $\lambda_{c,2}$ scale as $\Delta^2/\mu$ and are much smaller than $\lambda_c \sim \Delta$, at which the nodes disappear. The evolution at
 $\lambda \lesssim \lambda_c$
 is shown in Fig.~\ref{fig:2}. }
\label{fig:5}
\end{center}
\end{figure}

In this Appendix we discuss  the initial evolution of the
 NSL from radial beams directed transverse to the FSs, to tangential lines,
 directed along the FSs.  We present the results for circular pockets.  The evolution of the
 NSL for elliptical pockets is quite similar.

In the normal state, the dispersions for hybridized circular pockets are $\xi^{+-}_k = \xi_k \pm \lambda$, where $\xi_k = (k^2 - k^2_F)/(2m)$ is the dispersion in the absence of hybridization. In the superconducting state the dispersion is given by
Eq. (\ref{ch_1})
\begin{align}
\omega_{1,2}^2 ({\vec k}) =  ( \Delta^2 + y^2 \Delta^2 + \xi^2 + \lambda^2)
\mp 2 \sqrt{ S }   \, ,
\label{ch_1_1}
\end{align}
where $S = \xi^2 \lambda^2 +  y^2 \Delta^2 (\Delta^2 + \lambda^2) $ and $y = \alpha \cos 2\theta_k$.
At $\lambda =0$, Eq. (\ref{ch_1_1}) reduces to a conventional expression $\omega_{1,2}^2 ({\vec k}) = \xi^2 + \Delta^2 (1 \pm y)^2 $.  The dispersions in the normal and superconducting state coincide along 4 radial lines specified by $y = \pm 1$. Along these directions, $s^{\pm}$ gap has accidental nodes on one or the other electron FS.

Once $\lambda$ becomes non-zero, the
lines transform into banana-type loops still elongated transverse to the FS, and ``domes'' at $k > k_F$ (Fig.~\ref{fig:5}a) The loops close up
at $k=0$, and  at $k=k_F$ and $y \approx \pm 1$.  The NSL then
 undergo several topological changes
 at $\lambda \sim \Delta^2/\mu$, which is much smaller than critical $\lambda_c \sim \Delta$.
  At the first critical
$\lambda = \lambda_{c,1}$ the bananas touch pairwise along the directions $\theta_k = 0, \pm \pi/2$ and $\pi$ (Fig.~\ref{fig:5}b).
 At larger $\lambda$,
  eight bananas transform into
   four configurations,
 which resemble rabbit ears (Figs.~\ref{fig:5} c-d). The value of $\lambda_{c,1}$ is
\be
\lambda_{c,1}
\approx
\frac{ \Delta^2 }{ 4 \mu }
\frac{\left(\alpha ^2-1\right)^2}{\alpha^2 + 1 } \, .
\ee
At the next critical
\be
\lambda_{c,2} = \frac{\Delta^2 }{ 4 \mu}
\ee
the ``rabbit ears'' touch
each other along another set of symmetry directions,
$\theta_k = \pm \pi/4, \pm 3 \pi / 4$ (Fig.~\ref{fig:5}e), and at
 $\lambda > \lambda_{c,2}$  detach from $k=0$ ((Fig.~\ref{fig:5}f).
The evolution of the NSL
at larger $\lambda \sim \Delta$ is discussed in the main text and is shown in
Fig.~\ref{fig:3}.

\section{Fitting of the ARPES data}
\label{sec:app_B}

\begin{figure}[h]
\begin{center}
\includegraphics[width=0.95\columnwidth]{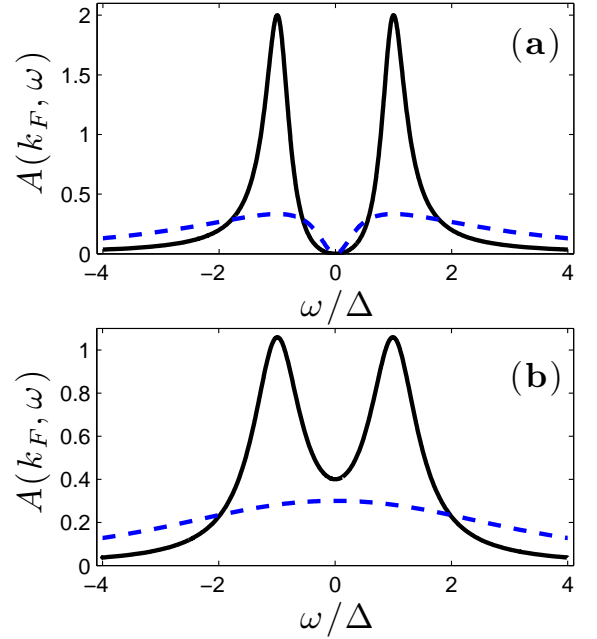}
\caption{Color online.
The spectral function $A(k_F, \omega)$ (a.u.), Eq.~\eqref{B1},  with self-energy
 given by (a) Eq.~\eqref{B2} and (b) Eq.~\eqref{B3}.
The solid black line is for $\gamma = 0.5 \Delta$.
The dashed blue line is  for $\gamma = 3.0 \Delta$.
 In the
case (a), the spectral function
 has a dip at $\omega=0$ and a peak at a finite frequency for  any non-zero $\Delta$.
In contrast, in the
case (b), the peak in the spectral function can be at $\omega =0$ despite that the gap is non-zero.
 This happens when the
  broadening is large enough, $\gamma > \sqrt{3} \Delta$. Notice that in this situation the peak at $\omega = 0$ is
   very broad.}
\label{fig:fit}
\end{center}
\end{figure}

In this Appendix we discuss the fitting procedure used in Ref.~\cite{recent_arpes} to extract the gap structure from ARPES data.
The authors of this work symmetrized the measured photoemission intensity to get rid of Fermi function and extract the spectral function $A(k, \omega) = (-1/\pi) G^{''}(k_F, \omega)$,  and used the convention that
the gap is finite when $A(k,\omega)$ at $k=k_F$ has two resolved maxima at finite frequencies, and zero when the maximum in the
spectral function is at zero frequency.  The same procedure was used earlier to identify
 Fermi ``arcs''
in the cuprates~\cite{arcs}.
This fitting procedure requires  care because the maximum in $A(k_F, \omega)$ can be at zero frequency even when the gap is non-zero.
This happens when fermionic damping is finite (as it always is at a finite $T$, even in a conventional s-wave superconductor)
 and is larger than the gap.

To extract the gap from the data, the authors of \cite{recent_arpes} related the spectral function to fermionic self-energy in a standard way, as
 \be
A(k_F,\omega) = - \frac{1}{\pi} \frac{\Sigma^{''} (k_F,\omega)}{(\omega - \Sigma^{'} (k_F, \omega)^2 + (\Sigma^{''} (k_F,\omega))^2}
\label{B1}
 \ee
and modeled the self-energy at each $k_z$ as
\be
\Sigma (k_F, \omega) = - i \gamma + \frac{\Delta^2}{\omega}\, .
\label{B2}
\ee
If one uses this fit, one finds that the maximum in the spectral function is at $\omega =0$ {\it only} when the superconducting gap
vanishes (see Fig~\ref{fig:fit}a). Indeed, the self-energy
given by \eqref{B2} diverges at $\omega = 0$ and therefore the spectral density has a dip at $\omega = 0$ for any
 nonzero  $\Delta$.
 Because the
measured spectral function $\alpha$ hole pocket  is peaked at $\omega =0$ near $k_z = \pi$,
the authors of \cite{recent_arpes} concluded that the gap on the $\alpha$ pocket must vanish for this $k_z$.

We argue that the fermionic self-energy of a dirty BCS superconductor
 contains the damping term $i\gamma$ not only as the stand-alone constant but also in the denominator of the $\Delta^2$ term, along with $\omega$, i.e., the true self-energy is
\be
\Sigma_t (k_F, \omega) = - i \gamma + \frac{\Delta^2}{\omega+i \gamma} \, .
\label{B3}
\ee
This form of the self-energy has been extensively discussed in the context of the physics of Fermi arcs in the cuprates~\cite{norman,millis,we}.
Using this self-energy, one obtains that the spectral function has a maximum at $\omega =0$ even when the gap is finite,
  provided that $\gamma > \sqrt{3} \Delta$.
We show in Fig.~\ref{fig:fit}b the spectral function obtained
 using the self-energy from Eq.~\eqref{B3} with a non-zero $\Delta$.  We see that the maximum in $A(k_F, \omega)$ is either at a finite frequency or at $\omega=0$, depending on the interplay between $\Delta$ and $\gamma$.  Observe that, when the  maximum is at $\omega =0$,  it is quite broad, much like in the experimental data in ~\cite{recent_arpes} near $k_z =\pi$.  We argue therefore that the data of the ARPES study in~\cite{recent_arpes}
 are in fact consistent with anisotropic but still no-nodal gap along the $\alpha$ pocket.
\end{appendix}

\end{document}